\documentclass{osa-article}
\usepackage{multirow}
\journal{oe}

\articletype{Research Article}

\begin{document}

\title{Ghost imaging based on Y-net: a dynamic coding and conjugate-decoding approach} 
\author{Ruiguo Zhu,\authormark{1,2} Hong Yu,\authormark{1,3,*} Zhijie Tan,\authormark{1,2}  Ronghua Lu,\authormark{1} ShenSheng Han,\authormark{1,3} Zengfeng Huang,\authormark{4} and Jian Wang\authormark{4}}


\address{\authormark{1}Key Laboratory for Quantum Optics and Center for Cold Atom Physics of CAS, Shanghai Institute of Optics
and Fine Mechanics, Chinese Academy of Sciences, Shanghai, 201800, China}
\address{\authormark{2}Center of Materials Science and Optoelectronics Engineering, University of Chinese Academy of Sciences, Beijing 10049, China}
\address{\authormark{3}Hangzhou Institute for Advanced Study, University of Chinese Academy of Sciences, Hangzhou 310024}
\address{\authormark{4}School of Data Science and Fudan-Xinzailing Joint Research Centre for Big Data, Fudan University Shanghai 200433, China}

\email{\authormark{*}yuhong@siom.ac.cn}



\begin{abstract}
Ghost imaging incorporating deep learning technology has recently attracted much attention in the optical imaging field. However, deterministic illumination and multiple exposure are still essential in most scenarios. Here we propose a ghost imaging scheme based on a novel conjugate-decoding deep learning framework (Y-net), which works well under both deterministic and indeterministic illumination. Benefited from the end-to-end characteristic of our network, the image of a sample can be achieved directly from a pair of correlated speckles collected by the detectors, and the sample is illuminated only once in the experiment. The spatial distribution of the speckles encoding the sample in the experiment can be completely different from that of the simulation speckles for training, as long as the statistical characteristics of the speckles remain unchanged. This approach is particularly important to high-resolution x-ray ghost imaging applications due to its potential for improving image quality and reducing radiation damage. And the idea of conjugate-decoding network may also be applied to other learning-based imaging.
\end{abstract}


\section{Introduction}
Ghost imaging (GI) extracts the information of an object by measuring the intensity correlation of optical fields, which has now has been widely applied in remote sensing, super-resolution, x-ray imaging, atoms and electron imaging etc.\cite{PhysRevLett.96.063602,padgett2017introduction,meyers2011turbulence,Backscattering_Bian,PhysRevA.92.013823,Erkmen:12,yu2016fourier,pelliccia2016experimental,khakimov2016ghost,schneider2018quantum,zhang2018tabletop,kingston2018ghost,li2018electron}.
In the traditional GI, a large number of measurements are required to calculate the ensemble average as an unbiased estimation of the sample's image, which brings a heavy burden to the imaging system. Later on, the compressive sensing framework has been combined with GI schemes, and the image quality has been greatly improved by exploiting the sparse prior of objects\cite{katz2009compressive,zerom2011entangled,yu2014adaptive,liu2016spectral,yu2015structured,zhu2018spatial}. To improve the sampling efficiency, some researches have taken advantage of the coding theory and designed the sensing matrix by optimizing the illuminating optical fields\cite{katkovnik2012compressive,Hu:19}. In the meantime, computational ghost imaging has emerged\cite{shapiro2008computational,hardy2013computational,sun20133d}. It is a deterministic measuring process, in which the incident light is preset or prerecorded.
Recently, deep learning techniques have been introduced into computational ghost imaging and the measurement rate goes down to a cheerful level that is comparable with compressive sensing and even lower\cite{lyu2017deep,wang2019learning,shimobaba2018computational}. In their work, the illuminating speckles encoding the sample are the same during the training and imaging process. However, in many GI scenarios, such as high-resolution x-ray ghost imaging, particle ghost imaging, and some remote sensing applications, the intensity distribution of the illumination fluctuates randomly and is difficult to be precisely manipulated\cite{khakimov2016ghost,yu2016fourier,li2018electron,schneider2018quantum,kingston2018ghost}.
\par
As a widespread machine learning framework, deep learning has demonstrated its magic power in many fields. In the literature of optical imaging, the deep learning inspired approaches have been demonstrated in compressive sensing\cite{bora2017compressed,quan2018compressed}, scatter imaging\cite{horisaki2016learning,sinha2017lensless,sun2018efficient,li2018imaging,li2018deep}, super-resolution\cite{dong2014learning,ledig2017photo}, microscopy\cite{thanh2018deep,nehme2018deep} and phase retrieval\cite{rivenson2018phase,goy2018low}.  
The imaging problems can be expressed as an optimization process
\begin{equation}\label{eqs:opt}
\min_{\mathbf{x}} {\|\Theta(\mathbf{x})-\mathbf{y}\|_2
+\gamma \Phi (\mathbf{x})},
\end{equation}
where $\mathbf{x}$ is the signal, $\mathbf{y}$ is the measurement, $\|\cdot\|_2$ is the $\ell_2$ norm, $\Theta$ is the forward operation, $\Phi$ and $\gamma$ are the regularization operation and its weight factor respectively.
A typical deep learning strategy is  to learn the representation $\Psi$ of the signal  $\mathbf{x}$ from the data set $\{\mathbf{x}\}$ and then optimize the latent variable  ${\mathbf{z}}$ \cite{bora2017compressed}. 
Note that $\mathbf{x} = \Psi (\mathbf{z})$ and the latent variable $\mathbf{z}$ usually belong to a lower dimension space, so the optimization is more convenient than finding $\mathbf{x}$ directly. 
Another strategy is to find the inverse operation of the forward operation $\Theta$ and the regularization operation $\Phi$ though the data set $\{\mathbf{x},\mathbf{y}\}$\cite{horisaki2016learning,sinha2017lensless,sun2018efficient,rivenson2018phase}.
Physics-informed priors are important for generalization during the training phase\cite{icsil2019deep,goy2018low}. In this strategy, the learned map $f \colon \mathbf{y} \mapsto \mathbf{x}$ is  directly related to the measurement $\mathbf{y}$, so this end-to-end map relies on a specifically determined $\Theta$. 
The problem becomes more challenging for a dynamical system where the forward operation $\Theta$ is not deterministic. Some researches focused on using the memory effect of scattering medium to characterize the statistical similarity which is invariant in scattering imaging\cite{freund1988memory,katz2014non}. Li obtained the network map using a set of fixed diffusers after a long time of training data acquisition\cite{li2018deep}, and multiple scattering images are inevitably necessary to capture sufficient statistical variations. 
\par
In this paper, we propose a conjugate-decoding deep learning framework (Y-net) for dynamic GI systems, which means the intensity distribution of the illuminating light in the system is indeterministic. A ghost imaging scheme based on the Y-net has been demonstrated. The network is trained with simulation data, and the testing results show that it has strong generalization capability and works well in the experiment. In our scheme, only a pair of correlated speckle patterns is needed to reconstruct the image of a sample, and the sample is illuminated only once. Thus, it provides potential application in x-ray imaging, in which the exposure should be reduced as much as possible considering the radiation damage of samples. In addition, our approach is based on an end-to-end network, so that the image of a sample can be directly obtained from the data collected by the detectors without extra processing, such as initial input image calculation, subsequent phase recovery, etc.

\section{Methods}\label{method}
\subsection{Imaging scheme }
\begin{figure}[ht]
\centering
\fbox{\includegraphics[width = 0.6\linewidth]{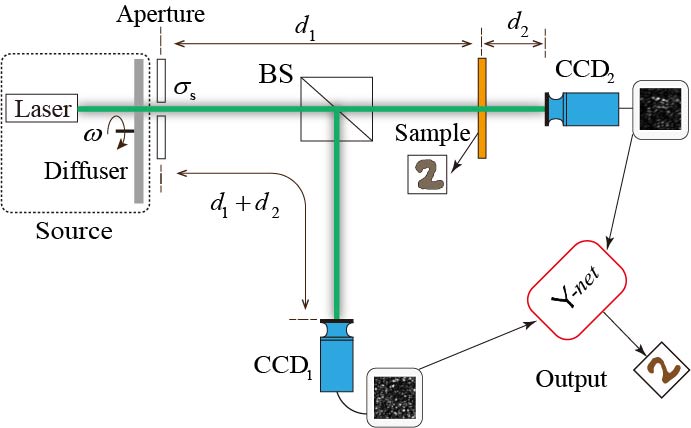}}
\caption{Experimental scheme of a ghost imaging system based on  Y-net. A pair of speckle images acquired by the two $\rm CCD$s are transferred to a well trained Y-net as input, and the output of the network is the image of the sample.}
\label{fig: setup_figure}
\end{figure}

The experimental scheme of ghost imaging based on Y-net is shown in Fig. \ref{fig: setup_figure}.
The laser illuminates a rotating ground glass to produce pseudo-thermal light. A diaphragm behind the ground glass is used to control the source size $\sigma_s$. A beam splitter divides the incident light into two beams: a reference beam propagating directly to the detector and a test beam with the sample inserted in the optical path. In the test beam, the distance from source to sample and the distance from sample to $\rm CCD$ are $d_1$ and $d_2$, respectively. In the reference beam, the distance from source to $\rm CCD$ is $d = d_1 + d_2$. A pair of speckle images acquired by the two $\rm \rm CCD$s are transferred to a well trained Y-net as the input, and the output of the network is the image of the sample.
\par 
The speckle distribution recorded by $\rm CCD_1$ can be described as following
\begin{equation}\label{Ir}
I_r(x_r) = \left|\int_{\sigma_s}E_0(x_0)h_d(x_0, x_r)dx_0 \right|^2,
\end{equation}
where $E_0(x_0)$ represents the optical field on the source plane, $h_d$ is the free-space transfer function from source to $ \rm CCD_1$.
The speckle distribution recorded by $\rm CCD_2$ is 
\begin{equation}\label{It}
I_t(x_t) = \left| \iint_{\sigma_s, {object}}E_0(x_0)h_{d_1}(x_0, x^\prime)t(x^\prime)h_{d_2}(x^\prime, x_t)dx_0dx^\prime \right |^2,
\end{equation}
where $t(x^\prime)$ denotes the transmittance of the sample, $h_{d_1}$ and $h_{d_2}$ are the free-space transfer functions from source to sample and from sample to $\rm CCD_2$, respectively.
Here the free-space transfer function between planes $x_1$ and $x_2$  is 
\begin{equation}\label{psf}
h_z(x_1,x_2) = \frac{e^{ikz}}{i\lambda z}\exp\left\{\frac{ik}{2z}\left(x_1-x_2\right)^2\right\},
\end{equation}
where $z$ is the distance between the two planes, $\lambda$ is the wavelength of the light and $k = \frac{2 \pi}{\lambda}$ .
\par
In traditional Fourier-transform ghost imaging (FGI), lots of measurements are required, and an ensemble average operation $\langle \cdot \rangle$ is used to obtain the Fourier-transform pattern of the sample, which is\cite{cheng2004incoherent,yu2016fourier}
\begin{equation}\label{gi}
\langle \Delta I_r(x_r)\Delta I_t(x_t)\rangle \propto  \left |T\left(\frac{x_r - x_t}{\lambda d_2}\right)\right|^2,
\end{equation}
where $T$ is the Fourier transformation of $t(x^\prime)$, and $\Delta I_k(x_k) = I_k(x_k) - \langle I_k(x_k) \rangle $, in which $k=r,t$. While FGI is combined with compressive sensing, a sensing equation is established according to the relationship between the speckle fields of the two beams. The relationship can be described as\cite{zhu2018spatial}
\begin{equation}\label{simply}
I_t(x_t) \propto \int_{ref} I_r(x_r) \left|T\left(\frac{x_r-x_t}{\lambda d_2}\right)\right|^2dx_r.
\end{equation}It can be discretized into a linear sensing equation
\begin{equation}\label{eqs:leq}
\mathbf{y} = \mathbf{Ab},
\end{equation}
in which $\mathbf{b}$ is the Fourier-transform pattern of the sample, $\mathbf{A}$ and $\mathbf{y}$ correspond to $I_r$ and $I_t$, respectively. The Fourier-transform pattern of the sample can be obtained by solving this equation. Then, a phase retrieval process needs to be carried out to recover the image of the sample from the Fourier-transform pattern in both traditional FGI and FGI combined with compressive sensing.
\par
In our Y-net based GI scheme, the two steps of Fourier-transform pattern acquisition and phase retrieval are integrated. The imaging problem can be modeled as Eq. (\ref{eqs:opt}) where the forward operation $\Theta$ is a composite operation $\mathbf{A}|\mathbf{T}(\cdot)|^2$. Here we use $\mathbf{x}$ represents the image of the object, $\mathbf{T}$ denotes the Fourier-transform matrix, and $|\cdot|^2$ is the point-wise square of modulus. The speckles in the reference beam $I_r$ is randomly distributed and indeterministic, so are the operation $\mathbf{A}$ and the composite operation $\mathbf{A}|\mathbf{T}(\cdot)|^2$. 
To  describe this indeternisitic system, we modify the model in Eq. (\ref{eqs:opt}) by adding an extra penalty concerning the dynamic measuring process, then the imaging problem can be expressed as
\begin{equation}\label{eqs:opt_dynamic}
\min_\mathbf{x} {\|\Theta(\mathbf{x})-\mathbf{y}\|_2+\alpha \|\Pi(\Theta)- \mathbf{A} \|_2
+\gamma \Phi (\mathbf{x})},
\end{equation}
where $\Pi$ represents the relationship between the forward operation $\Theta$ and the operation $\mathbf{A}$ which corresponds to the speckle distribution $I_r$, $\alpha$ is the corresponding weight factor.
\par
We solve this problem under the framework of deep learning. Instead of optimizing the signal $\mathbf{x}$, we optimize the model parameters by exploiting the training data and try to establish a direct map $f \colon \mathbf{y} \mapsto \mathbf{x}$ to obtain the image of a sample directly from measurement. The optimization is subject to the network parameters $\Omega^{-1} = \{\Theta^{-1}, \Pi^{-1}, \Phi^{-1}\}$ and we have
\begin{equation}\label{eqs:opt_dy}
\min_{\Omega^{-1}} {\|\Theta^{-1}(\mathbf{y})-\mathbf{x}\|_2+\alpha \|\Pi^{-1}(\mathbf{A})- \Theta \|_2
+\gamma \Phi^{-1} }.
\end{equation}
There are three types of parameters should be learned from data: (\romannumeral 1)   forward operation $\Theta$ which depends on $\mathbf{x}$ and $\mathbf{y}$; (\romannumeral 2 ) model transform $\Pi$ which depends on $A$ and $\Theta$; (\romannumeral 3) regularization $\Phi$ which depends on $\mathbf{x}$ and network architecture. Previous networks for GI\cite{lyu2017deep,shimobaba2018computational,wang2019learning}  just learnt the parameters (\romannumeral 1) and (\romannumeral 3). So their methods will be only suitable for deterministic (or static) systems, while our method can be implemented in both static and dynamic situations.  Besides, benefited from the end-to-end characteristic of our network, the sample's image can be reconstructed directly from the speckle images recorded by the $\rm CCD$s without subsequent phase retrieval requirement.

\subsection{Network architecture and training}\label{arch}
\begin{figure}[htp]
\centering
\includegraphics[width = 0.56\linewidth,height= 10.5 cm]{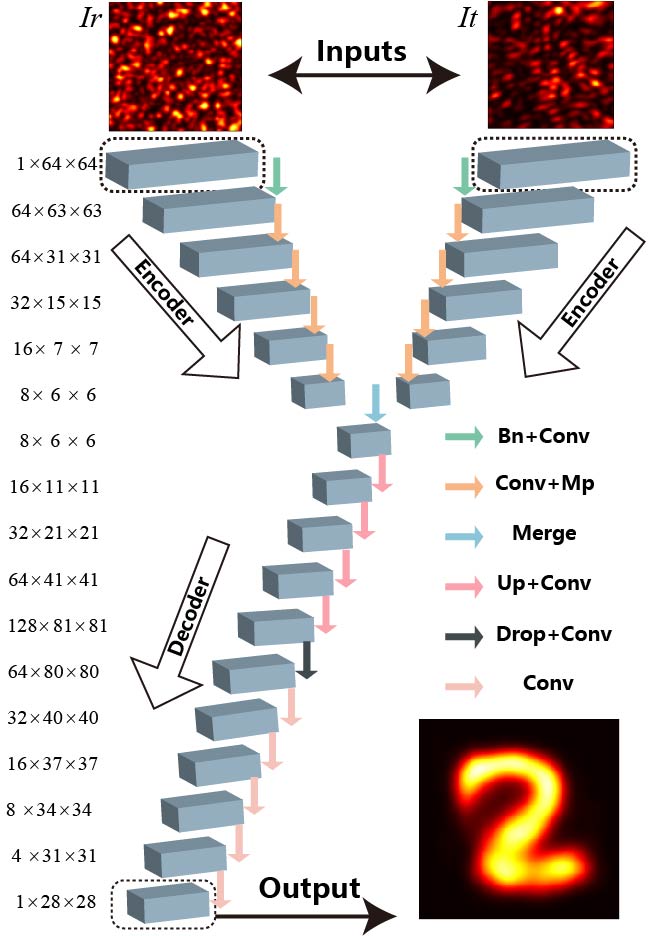}
\caption{ Architecture of the proposed Y-net. It consists of two encoders and one decoder. The input of the network is a pair of speckle distributions, and the output of the network is the image of the sample.}
\label{fig: net}
\end{figure}     
The overall structure of our Y-net consists of two encoders and one decoder as shown in Fig. \ref{fig: net}. The speckle patterns recorded by the detectors in the reference beam and the test beam are input into the two encoders separately in a symmetric way. Each encoder is composed of five convolutional layers with a batch normalization layer before the first convolutional layer and a max-pooling layer after each of the other four convolutional layers. Then the two encoders are merged by subtraction, and a decoder is built to recover the signal. The decoder has four upsampling layers, a dropout layer, and ten convolutional layers. More specifically, the decoder path firstly goes through four upsampling layers, each of which is followed by a convolutional layer, then passes a dropout layer followed by a convolutional layer, and next is a convolutional layer with a stride size of two, and finally through four convolutional layers with zero padding mode successively. The max-pooling size and the unsamping size are $2$. Without special explanation, all the convolutional filters have a size of $4\times 4$ and the padding mode is $1$. All the layers are followed by a rectified linear unit operation serving as an activation function, except for the last layer handled by a sigmoid function to restrict the range of pixels. 
\par  
This dual-encoder network is designed to extract the image information of an sample from the conjugate speckles recorded by the two detectors. From the perspective of coding theory, the optical measurement process can be regarded as the encoding part in our imaging scheme. The information of samples is encoded in the speckles detected in the test beam, while the original speckles are observed by the detector in the reference beam.  In the training process, our Y-net learns the encoding protocol of the optical system from a large number of correlated speckle pairs, and achieves the capability of decoding sample information directly from the raw conjugate speckle data. Thus, after training, the network serves as the decoding part of the imaging system. This is particularly useful when the encoding process of the imaging system changes dynamically.  
\par
It is expensive and time-consuming to acquire experiment data for training. As an alternative, the network is trained with simulation data.
The MNIST database was adopted to generate the training data.  For each digit image in the data set, we normalized its pixel values to the range of 0 to 1, and put it into the simulated GI system as a sample. The sample size is $1\times1$ $mm^2$ with a dimension of $28 \times 28$. The speckles in the reference beam and the test beam are synthesized according to Eqs. (\ref{Ir})(\ref{It})(\ref{psf}).  In each measurement, the optical field on the source plane is generated with different random phase distribution. The distance parameters $d_1$, $d_2$, and the diameter of the source $\sigma_s$ used in the simulation are the same as those in the experiment. The dimension of the detectors in the simulation is $64 \times 64$, and the pixel size is 46.88 ($5.86\times 8$) $\mu m$. The simulation data set was obtained in about 1 minute, and it included 70000 pairs of speckle patterns corresponding to 70000 digits in the MNIST database. 60000 pairs of the speckle patterns were used for training, and the remaining 10000 pairs were used for validation. 
    \par
    There are many types of loss functions can be chosen to train the network. In our work, we use the average binary cross-entropy as the loss function, which is defined by
    \begin{equation}
    L(P,Q) = -\frac{1}{2N}\sum_i^N[Q_i\log(P_i)+(1-Q_i)\log(1-P_i)],
    \label{eq:ace}
    \end{equation}
where $N$ is the pixel number, $Q_i$ and $P_i$ are the pixel value of target $Q$ and output $P$, respectively. The Adam optimizer is used to update the parameters with the initial learning rate $r = 0.002$ and $\beta_1 = 0.9$, $\beta_2 = 0.99$. The total training epoch is 250, and the training process took about 14 hours. After training, the experiment result of a sample can be given within several milliseconds. All computations including training and evaluation were performed on a workstation(@Intel-Xeon CPU and 4$\times$@Nvidia-GeForce-1080Ti GPUs).
\section{Results  and discussions} 
A 532 nm laser was adopted in the experiment, the distance parameters were $d_1= 5$ cm and $d_2 = 20.1$ cm, and the source diameter was $\sigma_s = 1$ mm.
The pixel size of the $\rm CCD$s was $5.86\times 5.86$ $\mu m^2$ and the number of pixels was $512\times 512$.
The speckle patterns recorded by the $\rm CCD$s were merged into $64\times 64$ and normalized before being transferred to the well-trained network. 
\begin{figure}[htbp]
\centering
\includegraphics[width =0.45\linewidth]{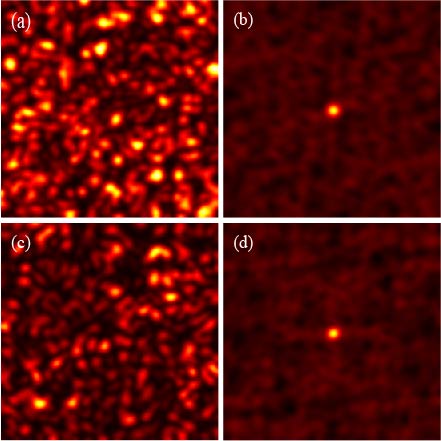}
\caption{Comparison between the speckles collected in the experiment and generated by simulation. (a) and (c) are the speckle images recorded by $\rm CCD_1$ and generated by simulation respectively. (b) and (d) are the corresponding second-order auto-correlation of the speckle images in (a) and (c). }\label{fig:comparsion}
\end{figure}
We compared the speckle patterns collected in the experiment and generated by simulation. Figure \ref{fig:comparsion}(a) is a typical speckle image recorded by $\rm CCD_1$, and Fig. \ref{fig:comparsion}(c) is a speckle image generated by simulation in the reference beam as described in the training process. Obviously, the spatial distributions of the two images are different. We calculated the corresponding second-order auto-correlation of the two speckle images, and the results are shown in Fig. \ref{fig:comparsion}(b) and (d). It can be found that the statistical characteristics of the two speckle images are almost the same. This is why our network uses simulation data in training, but after training it can be applied to experimental data.  
\par
We tested five samples("1","2","4","6","9") fabricated with stainless steel in the experiment. They were chosen from the testing part of the MNIST database, and never appeared in the training process. 
\begin{figure}[htbp]
\centering
\includegraphics[width =0.45\linewidth]{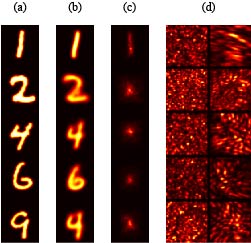}
\caption{ Experimental results. (a) is the original images of the samples, (b) is the output of our network, (c) is the results obtained by traditional ghost imaging method, (d) is the corresponding reference speckles (left) and test speckles (right).}
\label{fig: result_exp}
\end{figure}
The sample size was $1\times1$ $mm^2$.  
Figure \ref{fig: result_exp} presents the experimental results.  
Figure \ref{fig: result_exp}(b) gives the outcome of our network for the five samples. They are in good agreement with the original images in Fig. \ref{fig: result_exp}(a). For each sample, only one frame of reference speckle and one frame of test speckle were utilized to extract the image of the sample. 
As a comparison, we processed the speckle data using the traditional FGI method according to Eq. (\ref{gi}), in which all the pixels were used to calculate the ensemble average. And the hybrid input-output algorithm\cite{fienup1982phase} was adopted for phase retrieval. The reconstructed results are shown in Fig. \ref{fig: result_exp}(c). Unfortunately, there is almost no sign of digits in these images. 
The corresponding speckle data are displayed in Fig. \ref{fig: result_exp}(d).
It can be observed that the speckles are randomly distributed and changed dynamically. Thus, our method works well with this indeterministic GI system.
\begin{table}[ht]
\centering
\caption{\bf Quantitative evaluation of the image quality}
\begin{tabular}{|c|c|c|c|c|}
\hline
\multirow{2}{*}{Sample} & \multicolumn{2}{c|}{SSIM} & \multicolumn{2}{c|}{PSNR} \\ \cline{2-5} 
                        & Y-net     & GI         & Y-net     & GI        \\ \hline
"1"                     & 0.9134       & 0.2693     & 21.1719      & 11.5116    \\ \hline
"2"                    & 0.5809       & 0.0992     & 12.8016      & 6.7098     \\ \hline
"4"                     & 0.5701       & 0.1464     & 13.0317      & 8.0341     \\ \hline
"6"                     & 0.5797       & 0.1449     & 12.6473      & 9.2059     \\ \hline
"9"                     & 0.7046       & 0.1589     & 14.5564      & 8.1057     \\ \hline
\end{tabular}
 \label{tab:evaluation}
\end{table}
\par
To assess the quality of the image results, we used two evaluations: structural similarity index(SSIM) and peak signal-to-noise ratio(PSNR). The SSIM is defined by \cite{wang2004image}
\begin{equation}
SSIM(U,V) = \frac{(2\mu_u \mu_v + C_1)(2\sigma_{uv}+C_2)}{(\mu_u^2+ \mu_v^2 +C_1 )(\sigma_u^2+\sigma_v^2 +C_2)},
\end{equation}  
in which $U$ is the image to be evaluated, and V is the reference image, $\{\mu_u,\sigma_u\}$ and $\{\mu_v,\sigma_v\}$ are the means and variances of $U$ and $V$ respectively, $\sigma_{uv}$ is the co-variance of U and V, and ${C_1,C_2}$ are constants to prevent division by a small denominator. The PSNR is defined by
\begin{subequations}
\begin{eqnarray}
 &PSNR(U,V) = 10 \log_{10}\frac{MAX_I^2}{MSE(U,V)},\\
 &MSE(U,V) = \frac{1}{N}\sum_i^N(U_i - V_i)^2,
\end{eqnarray}
\end{subequations}
where $MAX_I$ is the maximum value of the image and $MAX_I = 1$ in this paper. The results of these two kinds of evaluations are summarized in Table \ref{tab:evaluation}. It is clear that our Y-net based GI method has better SSIM and PSNR. 
\par
\begin{figure}[ht]
\centering
\includegraphics[width =0.6\linewidth]{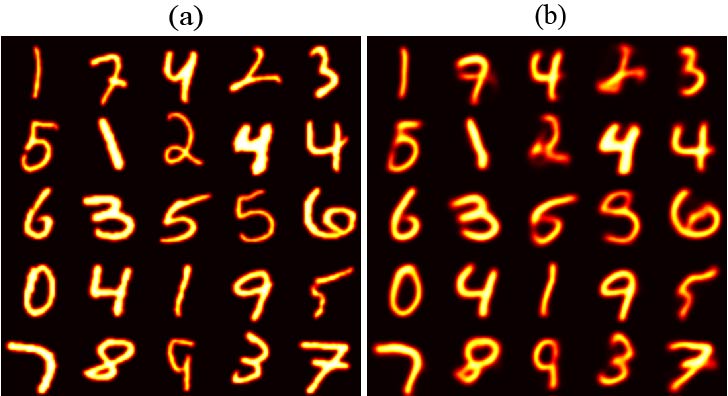}
\caption{Network performance for the static system. (a) is the original images, (b) is the output of the network. }\label{fig:static}
\end{figure}
\begin{figure}[hb]
\centering
\includegraphics[width =0.90\linewidth]{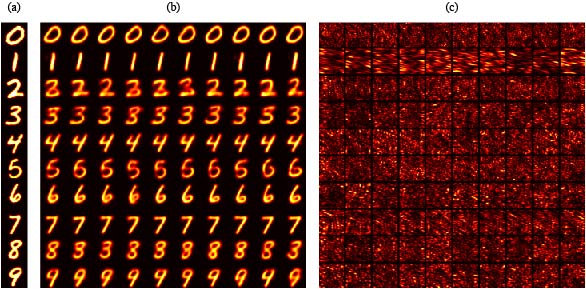}
\caption{Network performance with dynamic illumination. (a) is the original images, (b) is the network outputs, and (c) is the corresponding input speckles. For each digit sample, the simulation experiment was repeated 10 times. }
\label{fig: stability}
\end{figure}

We investigated the performance of our network in static and dynamic situations through more simulation experiments. In the static experiments, the imaging process remained unchanged, which means the forward operation was deterministic. To simulate the static GI system, we fixed random seeds for each sample when generating speckles, so that the reference speckles for each sample are identical. Based on the same network architecture described in subsection \ref{arch}, we trained the network again with the derived data set for the static GI system. Then we tested the network with a set of different digits shown in Fig. \ref{fig:static}(a), and Fig. \ref{fig:static}(b) presents the results. The images of samples are successfully obtained with high image quality.
In the dynamic experiments, we tested the stability of the network output when the illuminating speckles were generated randomly. We chose 10 digits from the testing part of the MNIST database.  For each digit sample, we repeated the experiment 10 times. Figure \ref{fig: stability} displays the results. Although the  input speckles are different as shown in Fig. \ref{fig: stability}(c), the network outputs presented in Fig. \ref{fig: stability}(b) are quite consistent with the original images in Fig. \ref{fig: stability}(a). It indicates that the our Y-net is stable and reliable for GI systems exploiting dynamical illumination. 
    \par
Many factors may affect the network performance, and the interaction among them is complicated. Even though, some methods are helpful to improve network accuracy, such as introducing physics-informed priors, adding functional layers, data enhancement, and so on.  In our work, the batch-normalization layer is necessary, and the dropout layer is the key to avoid over fitting. The dropout rate $p$ should be carefully chosen to guarantee the generalization of the network. We observed the impact of the dropout rate, and chose a dropout rate $p=0.6$.  

\section{Conclusion}
In summary, we have demonstrated a ghost imaging scheme based on Y-net, a novel conjugate-decoding deep learning framework that can be used to reconstruct sample images in both static and dynamic GI systems. As long as the statistical characteristics of the illuminating light remain unchanged, the image of a sample can be successfully achieved even if the spatial distribution of the illuminating light is indeterministic. Due to its strong generalization capability, Y-net can be applied to experimental data after training with simulation data. Thus, it can avoid the common difficulty of insufficient training data in learning-based imaging methods. Moreover, in previous GI schemes based on deep learning, the sample needs to be illuminated repeatedly to obtain enough measurements, and sometimes a set of sample images obtained by the traditional GI methods are required as input. But in our scheme, the sample will be illuminated only once, and benefited from the end-to-end characteristic of the Y-net, the sample image can be extracted directly from the conjugate speckle data collected by the detectors. Finally, compared with the traditional Fourier-transform GI techniques which involve an ill-posed phase retrieval problem, Y-net ghost imaging is more convenient, and can greatly improve the sampling efficiency and image quality. All these features are of great significance for GI applications, especially for imaging that requires dynamic illumination and single exposure of samples. It is particularly useful in high-resolution x-ray ghost imaging due to its potential for achieving high quality image with random speckles and reducing radiation damage. And the idea of conjugate-decoding network may also be applied to other imaging scenarios.
\section*{Funding}
National Natural Science Foundation of China (11627811); National Key Research and Development Program of China (2017YFB0503303, 2017YFB0503300).


\section*{Disclosures}
The authors declare no conflicts of interest.

\if false
\section{Multiple corresponding authors}

There are two options for indicating multiple corresponding authorship, and they are formatted quite differently. The first format would be as follows and uses an asterisk to denote one of the authors:

\begin{verbatim}
\author{Author One\authormark{1,3} and Author Two\authormark{2,4,*}}

\address{\authormark{1}Peer Review, Publications Department,
Optical Society of America, 2010 Massachusetts Avenue NW,
Washington, DC 20036, USA\\
\authormark{2}Publications Department, Optical Society of America,
2010 Massachusetts Avenue NW, Washington, DC 20036, USA\\
\authormark{3}xyz@osa.org}

\email{\authormark{*}opex@osa.org}
\end{verbatim}

This format will generate the following appearance:

\medskip

\author{Author One\authormark{1,3} and Author Two\authormark{2,4,*}}

\address{\authormark{1}Peer Review, Publications Department,
Optical Society of America, 2010 Massachusetts Avenue NW,
Washington, DC 20036, USA\\
\authormark{2}Publications Department, Optical Society of America,
2010 Massachusetts Avenue NW, Washington, DC 20036, USA\\
\authormark{3}xyz@osa.org}

\email{\authormark{*}opex@osa.org}

\medskip

The second format forgoes the asterisk and sets all email addresses equally within the affiliations. Please note that this format does not use the \verb+\email{}+ field at all.
\begin{verbatim}
\author{Author One\authormark{1,3} and Author Two\authormark{2,4}}

\address{\authormark{1}Peer Review, Publications Department,
Optical Society of America, 2010 Massachusetts Avenue NW,
Washington, DC 20036, USA\\
\authormark{2}Publications Department, Optical Society of America,
2010 Massachusetts Avenue NW, Washington, DC 20036, USA\\
\authormark{3}xyz@osa.org\\
\authormark{4}opex@osa.org}
\end{verbatim}

This format will generate the following appearance:

\medskip

\author{Author One\authormark{1,3} and Author Two\authormark{2,4}}

\address{\authormark{1}Peer Review, Publications Department,
Optical Society of America, 2010 Massachusetts Avenue NW, Washington, DC 20036, USA\\
\authormark{2}Publications Department, Optical Society of America, 2010 Massachusetts Avenue NW, Washington, DC 20036, USA\\
\authormark{3}xyz@osa.org\\
\authormark{4}opex@osa.org}
\medskip
These are the preferred
formats for multiple corresponding authorship, and either may be used.

\section{Abstract}
The abstract should be limited to approximately 100 words.
If the work of another author is cited in the abstract, that citation should be written out without a number, (e.g., journal, volume, first page, and year in square brackets [Opt. Express {\bfseries 22}, 1234 (2014)]), and a separate citation should be included in the body of the text. The first reference cited in the main text must be [1]. Do not include numbers, bullets, or lists inside the abstract.

\begin{figure}[h!]
\centering\includegraphics[width=7cm]{osafig1}
\caption{Sample caption (Fig. 2, \cite{Yelin:03}).}
\end{figure}

\section{Assessing final manuscript length}
OSA's Universal Manuscript Template is based on the OSA Express layout and will provide an accurate length estimate for Optics Express, Biomedical Optics Express,  Optical Materials Express, and OSA's newest title OSA Continuum. Applied Optics, JOSAA, JOSAB, Optics Letters, Optica, and Photonics Research publish articles in a two-column layout. To estimate the final page count in a two-column layout, multiply the manuscript page count (in increments of 1/4 page) by 60\%. For example, 11.5 pages in the OSA Universal Manuscript Template are roughly equivalent to 7 composed two-column pages. Note that the estimate is only an approximation, as treatment of figure sizing, equation display, and other aspects can vary greatly across manuscripts. Authors of Letters may use the legacy template for a more accurate length estimate.

\section{Figures, tables, and supplemental materials}

\subsection{Figures and tables}

OSA encourages authors to submit color figures with their manuscripts. Figures and tables should be placed in the body of the manuscript. Standard \LaTeX{} environments should be used to place tables and figures:
\begin{verbatim}
\begin{figure}[htbp]
\centering\includegraphics[width=7cm]{osafig1}
\caption{Sample caption (Fig. 2, \cite{Yelin:03}).}
\end{figure}
\end{verbatim}

\subsection{Supplementary materials in OSA journals}

OSA journals allow authors to include supplementary materials as integral parts of a manuscript. Such materials are subject to peer-review procedures along with the rest of the paper and should be uploaded and described using OSA's Prism manuscript system. Please see the \href{http://www.osapublishing.org/submit/style/multimedia.cfm}{Author Guidelines for Supplementary Materials in OSA Journals} for further information.

Supplementary materials must be associated with a figure, table, or equation, OR be referenced in the results section of the manuscript. Please note that to create text color for supplementary materials links, use of the command \\
\verb|\textcolor{urlblue}{Visualization 1}| is preferred to using the command\\
\verb|\url{Visualization 1}|.

\begin{figure}[ht!]
\centering\includegraphics{osafig2}
\caption{(a) Three traps create three rings of magnetic nanoparticles. (b) The rings interact with one another (see \textcolor{urlblue}{Visualization 1}, \cite{Masajada:13}).}
\end{figure}

\begin{verbatim}
\begin{figure}[hbt!]
\centering\includegraphics{opexfig2}
\caption{Normalized modulus distributions of transverse electrical
field components of the TM01 mode in PWs with (a) SiO_2 core
and (b) Si core}{Visualization 1}), \cite{Masajada:13}).}
\end{figure}
\end{verbatim}

\section{Mathematical and scientific notation}

\subsection{Displayed equations} Displayed equations should be centered.
Equation numbers should appear at the right-hand margin, in
parentheses:
\begin{equation}
J(\rho) =
 \frac{\gamma^2}{2} \; \sum_{k({\rm even}) = -\infty}^{\infty}
	\frac{(1 + k \tau)}{ \left[ (1 + k \tau)^2 + (\gamma  \rho)^2  \right]^{3/2} }.
\end{equation}

All equations should be numbered in the order in which they appear
and should be referenced  from within the main text as Eq. (1),
Eq. (2), and so on [or as inequality (1), etc., as appropriate].

\section*{Funding}
Please identify all appropriate funding sources by name and contract number. Funding information should be listed in a separate block preceding any acknowledgments. List only the funding agencies and any associated grants or project numbers, as shown in the example below:\\
\\
National Science Foundation (NSF) (1253236, 0868895, 1222301); Program 973 (2014AA014402); Natural National Science Foundation (NSFC) (123456).\\
\\
OSA participates in \href{https://www.crossref.org/fundingdata/}{Crossref's Funding Data}, a service that provides a standard way to report funding sources for published scholarly research. To ensure consistency, please enter any funding agencies and contract numbers from the Funding section in Prism during submission or revisions.

\section*{Acknowledgments}
Acknowledgments, if included, should appear at the end of the document. The section title should not be numbered.

\section*{Disclosures}

Disclosures should be listed in a separate nonnumbered section at the end of the manuscript. List the Disclosures codes identified on OSA's \href{http://www.osapublishing.org/submit/review/conflicts-interest-policy.cfm}{Conflict of Interest policy page}, as shown in the examples below:

\medskip

\noindent ABC: 123 Corporation (I,E,P), DEF: 456 Corporation (R,S). GHI: 789 Corporation (C).

\medskip

\noindent If there are no disclosures, then list ``The authors declare no conflicts of interest.''

\section{References}
\label{sec:refs}
Proper formatting of references is extremely important, not only for consistent appearance but also for accurate electronic tagging. Please follow the guidelines provided below on formatting, callouts, and use of Bib\TeX.

\subsection{Formatting reference items}
Each source must have its own reference number. Footnotes (notes at the bottom of text pages) are not used in OSA journals. References require all author names, full titles, and inclusive pagination. Examples of common reference types can be found on the  \href{http://www.osapublishing.org/submit/style/style_traditional_journals.cfm} {Author and Reviewer Resource Center}.

The commands \verb+\begin{thebibliography}{}+ and \verb+\end{thebibliography}+ format the section according to standard style, showing the title {\bfseries References}.  Use the \verb+\bibitem{label}+ command to start each reference.

\subsection{Formatting reference citations}
References should be numbered consecutively in the order in which they are referenced in the body of the paper. Set reference callouts with standard \verb+\cite{}+ command or set manually inside square brackets [1].

To reference multiple articles at once, simply use a comma to separate the reference labels, e.g. \verb+\cite{Yelin:03,Masajada:13,Zhang:14}+, produces \cite{Yelin:03,Masajada:13,Zhang:14}.

\subsection{Bib\TeX}
\label{sec:bibtex}
Bib\TeX{} may be used to create a file containing the references, whose contents (i.e., contents of \texttt{.bbl} file) can then be pasted into the bibliography section of the \texttt{.tex} file. A Bib\TeX{} style file, \texttt{osajnl.bst}, is provided.

If your manuscript already contains a manually formatted \verb|\begin{thebibliography}|... \verb|\end{thebibliography}| list, then delete the \texttt{latexmkrc} file from your submission files. However you should ensure that your manually-formatted reference list adheres to the OSA style accurately.

\section{Conclusion}
After proofreading the manuscript, compress your .tex manuscript file and all figures (which should be in EPS or PDF format) in a ZIP, TAR or TAR-GZIP package. All files must be referenced at the root level (e.g., file \texttt{figure-1.eps}, not \texttt{/myfigs/figure-1.eps}). If there are supplementary materials, the associated files should not be included in your manuscript archive but be uploaded separately through the Prism interface.


Add references with BibTeX or manually.
\cite{Zhang:14,OSA,FORSTER2007,Dean2006,testthesis,Yelin:03,Masajada:13,codeexample}

\fi
\bibliography{references}

\begin{thebibliography}{10}
\newcommand{\enquote}[1]{``#1''}

\bibitem{PhysRevLett.96.063602}
G.~Scarcelli, V.~Berardi, and Y.~Shih, \enquote{Can two-photon correlation of
  chaotic light be considered as correlation of intensity fluctuations?}
  {\protect\JournalTitle{Phys. Rev. Lett.}} \textbf{96}, 063602 (2006).

\bibitem{padgett2017introduction}
M.~J. Padgett and R.~W. Boyd, \enquote{An introduction to ghost imaging:
  quantum and classical,} {\protect\JournalTitle{Philosophical Transactions of
  the Royal Society A: Mathematical, Physical and Engineering Sciences}}
  \textbf{375}, 20160233 (2017).

\bibitem{meyers2011turbulence}
R.~E. Meyers, K.~S. Deacon, and Y.~Shih, \enquote{Turbulence-free ghost
  imaging,} {\protect\JournalTitle{Applied Physics Letters}} \textbf{98},
  111115 (2011).

\bibitem{Backscattering_Bian}
M.~Bina, D.~Magatti, M.~Molteni, A.~Gatti, L.~A. Lugiato, and F.~Ferri,
  \enquote{Backscattering differential ghost imaging in turbid media,}
  {\protect\JournalTitle{Phys. Rev. Lett.}} \textbf{110}, 083901 (2013).

\bibitem{PhysRevA.92.013823}
D.-J. Zhang, H.-G. Li, Q.-L. Zhao, S.~Wang, H.-B. Wang, J.~Xiong, and K.~Wang,
  \enquote{Wavelength-multiplexing ghost imaging,} {\protect\JournalTitle{Phys.
  Rev. A}} \textbf{92}, 013823 (2015).

\bibitem{Erkmen:12}
B.~I. Erkmen, \enquote{Computational ghost imaging for remote sensing,}
  {\protect\JournalTitle{J. Opt. Soc. Am. A}} \textbf{29}, 782--789 (2012).

\bibitem{yu2016fourier}
H.~Yu, R.~Lu, S.~Han, H.~Xie, G.~Du, T.~Xiao, and D.~Zhu,
  \enquote{Fourier-transform ghost imaging with hard x rays,}
  {\protect\JournalTitle{Phys. Rev. Lett.}} \textbf{117}, 113901 (2016).

\bibitem{pelliccia2016experimental}
D.~Pelliccia, A.~Rack, M.~Scheel, V.~Cantelli, and D.~M. Paganin,
  \enquote{Experimental x-ray ghost imaging,} {\protect\JournalTitle{Phys. Rev.
  Lett.}} \textbf{117}, 113902 (2016).

\bibitem{khakimov2016ghost}
R.~I. Khakimov, B.~Henson, D.~Shin, S.~Hodgman, R.~Dall, K.~Baldwin, and
  A.~Truscott, \enquote{Ghost imaging with atoms,}
  {\protect\JournalTitle{Nature}} \textbf{540}, 100 (2016).

\bibitem{schneider2018quantum}
R.~Schneider, T.~Mehringer, G.~Mercurio, L.~Wenthaus, A.~Classen, G.~Brenner,
  O.~Gorobtsov, A.~Benz, D.~Bhatti, L.~Bocklage \emph{et~al.}, \enquote{Quantum
  imaging with incoherently scattered light from a free-electron laser,}
  {\protect\JournalTitle{Nature Physics}} \textbf{14}, 126--129 (2018).

\bibitem{zhang2018tabletop}
A.-X. Zhang, Y.-H. He, L.-A. Wu, L.-M. Chen, and B.-B. Wang, \enquote{Tabletop
  x-ray ghost imaging with ultra-low radiation,}
  {\protect\JournalTitle{Optica}} \textbf{5}, 374--377 (2018).

\bibitem{kingston2018ghost}
A.~M. Kingston, D.~Pelliccia, A.~Rack, M.~P. Olbinado, Y.~Cheng, G.~R. Myers,
  and D.~M. Paganin, \enquote{Ghost tomography,}
  {\protect\JournalTitle{Optica}} \textbf{5}, 1516--1520 (2018).

\bibitem{li2018electron}
S.~Li, F.~Cropp, K.~Kabra, T.~Lane, G.~Wetzstein, P.~Musumeci, and D.~Ratner,
  \enquote{Electron ghost imaging,} {\protect\JournalTitle{Phys. Rev. Lett.}}
  \textbf{121}, 114801 (2018).

\bibitem{katz2009compressive}
O.~Katz, Y.~Bromberg, and Y.~Silberberg, \enquote{Compressive ghost imaging,}
  {\protect\JournalTitle{Applied Physics Letters}} \textbf{95}, 131110 (2009).

\bibitem{zerom2011entangled}
P.~Zerom, K.~W.~C. Chan, J.~C. Howell, and R.~W. Boyd,
  \enquote{Entangled-photon compressive ghost imaging,}
  {\protect\JournalTitle{Physical Review A}} \textbf{84}, 061804 (2011).

\bibitem{yu2014adaptive}
W.-K. Yu, M.-F. Li, X.-R. Yao, X.-F. Liu, L.-A. Wu, and G.-J. Zhai,
  \enquote{Adaptive compressive ghost imaging based on wavelet trees and sparse
  representation,} {\protect\JournalTitle{Optics Express}} \textbf{22},
  7133--7144 (2014).

\bibitem{liu2016spectral}
Z.~Liu, S.~Tan, J.~Wu, E.~Li, X.~Shen, and S.~Han, \enquote{Spectral camera
  based on ghost imaging via sparsity constraints,} {\protect\JournalTitle{Sci.
  Rep.}} \textbf{6}, 25718 (2016).

\bibitem{yu2015structured}
H.~Yu, E.~Li, W.~Gong, and S.~Han, \enquote{Structured image reconstruction for
  three-dimensional ghost imaging lidar,} {\protect\JournalTitle{Optics
  Express}} \textbf{23}, 14541--14551 (2015).

\bibitem{zhu2018spatial}
R.~Zhu, H.~Yu, R.~Lu, Z.~Tan, and S.~Han, \enquote{Spatial multiplexing
  reconstruction for fourier-transform ghost imaging via sparsity constraints,}
  {\protect\JournalTitle{Optics Express}} \textbf{26}, 2181--2190 (2018).

\bibitem{katkovnik2012compressive}
V.~Katkovnik and J.~Astola, \enquote{Compressive sensing computational ghost
  imaging,} {\protect\JournalTitle{J. Opt. Soc. Am. A}} \textbf{29}, 1556--1567
  (2012).

\bibitem{Hu:19}
C.~Hu, Z.~Tong, Z.~Liu, Z.~Huang, J.~Wang, and S.~Han, \enquote{Optimization of
  light fields in ghost imaging using dictionary learning,}
  {\protect\JournalTitle{Opt. Express}} \textbf{27}, 28734--28749 (2019).

\bibitem{shapiro2008computational}
J.~H. Shapiro, \enquote{Computational ghost imaging,}
  {\protect\JournalTitle{Physical Review A}} \textbf{78}, 061802 (2008).

\bibitem{hardy2013computational}
N.~D. Hardy and J.~H. Shapiro, \enquote{Computational ghost imaging versus
  imaging laser radar for three-dimensional imaging,}
  {\protect\JournalTitle{Physical Review A}} \textbf{87}, 023820 (2013).

\bibitem{sun20133d}
B.~Sun, M.~P. Edgar, R.~Bowman, L.~E. Vittert, S.~Welsh, A.~Bowman, and
  M.~Padgett, \enquote{3d computational imaging with single-pixel detectors,}
  {\protect\JournalTitle{Science}} \textbf{340}, 844--847 (2013).

\bibitem{lyu2017deep}
M.~Lyu, W.~Wang, H.~Wang, H.~Wang, G.~Li, N.~Chen, and G.~Situ,
  \enquote{Deep-learning-based ghost imaging,} {\protect\JournalTitle{Sci.
  Rep.}} \textbf{7}, 17865 (2017).

\bibitem{wang2019learning}
F.~Wang, H.~Wang, H.~Wang, G.~Li, and G.~Situ, \enquote{Learning from
  simulation: An end-to-end deep-learning approach for computational ghost
  imaging,} {\protect\JournalTitle{Optics Express}} \textbf{27}, 25560--25572
  (2019).

\bibitem{shimobaba2018computational}
T.~Shimobaba, Y.~Endo, T.~Nishitsuji, T.~Takahashi, Y.~Nagahama, S.~Hasegawa,
  M.~Sano, R.~Hirayama, T.~Kakue, A.~Shiraki, and T.~Ito,
  \enquote{Computational ghost imaging using deep learning,}
  {\protect\JournalTitle{Optics Communications}} \textbf{413}, 147--151 (2018).

\bibitem{bora2017compressed}
A.~Bora, A.~Jalal, E.~Price, and A.~G. Dimakis, \enquote{Compressed sensing
  using generative models,} in \emph{Proceedings of the 34th International
  Conference on Machine Learning - Volume 70,}  (JMLR.org, 2017), ICML’17, p.
  537–546.

\bibitem{quan2018compressed}
T.~M. Quan, T.~Nguyen-Duc, and W.-K. Jeong, \enquote{Compressed sensing mri
  reconstruction using a generative adversarial network with a cyclic loss,}
  {\protect\JournalTitle{IEEE Trans Med Imaging}} \textbf{37}, 1488--1497
  (2018).

\bibitem{horisaki2016learning}
R.~Horisaki, R.~Takagi, and J.~Tanida, \enquote{Learning-based imaging through
  scattering media,} {\protect\JournalTitle{Optics Express}} \textbf{24},
  13738--13743 (2016).

\bibitem{sinha2017lensless}
A.~Sinha, J.~Lee, S.~Li, and G.~Barbastathis, \enquote{Lensless computational
  imaging through deep learning,} {\protect\JournalTitle{Optica}} \textbf{4},
  1117--1125 (2017).

\bibitem{sun2018efficient}
Y.~Sun, Z.~Xia, and U.~S. Kamilov, \enquote{Efficient and accurate inversion of
  multiple scattering with deep learning,} {\protect\JournalTitle{Optics
  Express}} \textbf{26}, 14678--14688 (2018).

\bibitem{li2018imaging}
S.~Li, M.~Deng, J.~Lee, A.~Sinha, and G.~Barbastathis, \enquote{Imaging through
  glass diffusers using densely connected convolutional networks,}
  {\protect\JournalTitle{Optica}} \textbf{5}, 803--813 (2018).

\bibitem{li2018deep}
Y.~Li, Y.~Xue, and L.~Tian, \enquote{Deep speckle correlation: a deep learning
  approach toward scalable imaging through scattering media,}
  {\protect\JournalTitle{Optica}} \textbf{5}, 1181--1190 (2018).

\bibitem{dong2014learning}
C.~Dong, C.~C. Loy, K.~He, and X.~Tang, \enquote{Learning a deep convolutional
  network for image super-resolution,} in \emph{European conference on computer
  vision,}  (Springer, 2014), pp. 184--199.

\bibitem{ledig2017photo}
C.~Ledig, L.~Theis, F.~Huszar, J.~Caballero, A.~Cunningham, A.~Acosta,
  A.~Aitken, A.~Tejani, J.~Totz, Z.~Wang, and W.~Shi, \enquote{Photo-realistic
  single image super-resolution using a generative adversarial network,} in
  \emph{The IEEE Conference on Computer Vision and Pattern Recognition (CVPR),}
   (2017).

\bibitem{thanh2018deep}
N.~Thanh, Y.~Xue, Y.~Li, L.~Tian, and G.~Nehmetallah, \enquote{Deep learning
  approach to fourier ptychographic microscopy,} {\protect\JournalTitle{Optics
  Express}}  (2018).

\bibitem{nehme2018deep}
E.~Nehme, L.~E. Weiss, T.~Michaeli, and Y.~Shechtman, \enquote{Deep-storm:
  super-resolution single-molecule microscopy by deep learning,}
  {\protect\JournalTitle{Optica}} \textbf{5}, 458--464 (2018).

\bibitem{rivenson2018phase}
Y.~Rivenson, Y.~Zhang, H.~G{\"u}nayd{\i}n, D.~Teng, and A.~Ozcan,
  \enquote{Phase recovery and holographic image reconstruction using deep
  learning in neural networks,} {\protect\JournalTitle{Light: Science \&
  Applications}} \textbf{7}, 17141 (2018).

\bibitem{goy2018low}
A.~Goy, K.~Arthur, S.~Li, and G.~Barbastathis, \enquote{Low photon count phase
  retrieval using deep learning,} {\protect\JournalTitle{Phys. Rev. Lett.}}
  \textbf{121}, 243902 (2018).

\bibitem{icsil2019deep}
C.~I{\c{s}}il, F.~S. Oktem, and A.~Ko{\c{c}}, \enquote{Deep learning-based
  hybrid approach for phase retrieval,} in \emph{Computational Optical Sensing
  and Imaging,}  (Optical Society of America, 2019), pp. CTh2C--5.

\bibitem{freund1988memory}
I.~Freund, M.~Rosenbluh, and S.~Feng, \enquote{Memory effects in propagation of
  optical waves through disordered media,} {\protect\JournalTitle{Phys. Rev.
  Lett.}} \textbf{61}, 2328 (1988).

\bibitem{katz2014non}
O.~Katz, P.~Heidmann, M.~Fink, and S.~Gigan, \enquote{Non-invasive single-shot
  imaging through scattering layers and around corners via speckle
  correlations,} {\protect\JournalTitle{Nature photonics}} \textbf{8}, 784
  (2014).

\bibitem{cheng2004incoherent}
J.~Cheng and S.~Han, \enquote{Incoherent coincidence imaging and its
  applicability in x-ray diffraction,} {\protect\JournalTitle{Phys. Rev.
  Lett.}} \textbf{92}, 093903 (2004).

\bibitem{fienup1982phase}
J.~R. Fienup, \enquote{Phase retrieval algorithms: a comparison,}
  {\protect\JournalTitle{Applied Optics}} \textbf{21}, 2758--2769 (1982).

\bibitem{wang2004image}
Z.~Wang, A.~C. Bovik, H.~R. Sheikh, and E.~P. Simoncelli, \enquote{Image
  quality assessment: from error visibility to structural similarity,}
  {\protect\JournalTitle{IEEE Trans. Image Process}} \textbf{13}, 600--612
  (2004).

\end{thebibliography}






\end{document}